\documentclass[twocolumn,showpacs,aps,prl,superscriptaddress]{revtex4}

\newcommand{\beq}{\begin{eqnarray}}
\newcommand{\eeq}{\end{eqnarray}}
\newcommand{\beqq}{\begin{eqnarray*}}
\newcommand{\eeqq}{\end{eqnarray*}}

\usepackage{graphicx}
\usepackage{dcolumn}
\usepackage{bm}
\usepackage{amsmath,amssymb,amsfonts}
\usepackage{color}

\begin{document}

\title{Theory of interacting topological crystalline insulators}
\author{Hiroki Isobe}
\affiliation{Department of Applied Physics, University of Tokyo, Bunkyo, Tokyo 113-8656, Japan}
\author{Liang Fu}
\affiliation{Department of Physics, Massachusetts Institute of Technology,
Cambridge, MA 02139, USA}

\begin{abstract}
We study the effect of electron interactions in topological crystalline insulators (TCIs) protected by mirror symmetry, which are realized in the SnTe material class and host multi-valley Dirac fermion surface states.   
We find that interactions reduce the integer classification of noninteracting TCIs in three dimensions, indexed by the mirror Chern number, to a finite group $Z_8$.  
In particular, we explicitly construct a microscopic interaction Hamiltonian to gap 8 flavors of Dirac fermions on the TCI surface, while preserving the mirror symmetry. 
Our construction builds on interacting edge states of $U(1)\times Z_2$ symmetry-protected topological (SPT) phases of fermions in two dimensions, which we classify. 
Our work reveals a  deep connection between 3D topological phases protected by spatial symmetries and 2D topological phases protected by internal symmetries.
\end{abstract}


\maketitle

The prediction and observation of topological crystalline insulators (TCIs) in the SnTe material class has expanded the scope of topological matter and gained wide interest~\cite{hsieh, ando, poland, hasan, AndoFu}. These TCIs possess topological surface states that are protected by mirror symmetry of the rocksalt crystal and become gapped under symmetry-breaking structural distortions~\cite{vidya1, vidya2, arpes, serbyn}. 
These surface states are predicted to exhibit a plethora of novel phenomena ranging from large quantum anomalous Hall conductance~\cite{hsieh, bernevig,zhang} to strain-induced pseudo-Landau levels and superconductivity~\cite{TangFu}, which are currently under intensive study~\cite{cha,gu, vidya3}.     

According to band theory, TCIs protected by mirror symmetry are classified by an integer topological invariant, the mirror Chern number~\cite{teofukane}. 
However, recent theoretical breakthroughs~\cite{kitaev,ryu, yao, qi, vishwanath, wang1,levin, coupledwire} have found that the classifications of interacting systems are markedly different from noninteracting systems in various classes of topological insulators and superconductors protected by internal symmetries~\cite{schnyder}.  
This raises the open question about the classification of interacting TCIs protected by spatial symmetries. On the experimental side, a growing body of interaction-driven phenomena has been found in existing TCI materials, including spontaneous surface structural transition and gap generation~\cite{vidya1, vidya2, arpes} and anomalous bulk band inversion~\cite{inversion}. Moreover, new TCI materials have been predicted in transition metal oxides~\cite{fiete, liuhsiehfu} and heavy fermion compounds~\cite{dai, sun}, where strong electron interactions are expected. 

Motivated by these theoretical and experimental developments, in this work we study the effect of electron interactions in mirror-symmetric TCIs.  
Our main result is that interactions reduce the classification of 3D TCIs from $Z$ in the noninteracting case to $Z_8$.     
We obtain this result by introducing a ``domain wall'' construction of interacting surface states of 3D TCIs, which exploits the nonlocal nature of mirror symmetry. 
This construction builds on interacting edge states of 2D TCIs or $U(1)\times Z_2$ symmetry-protected topological (SPT) phases, which we classify. 
Our work reveals a  deep connection between 3D topological phases protected by spatial symmetries and 2D topological phases protected by internal symmetries. 

{\bf Interacting TCIs in two dimensions:} We first study interacting TCIs in two dimensions to set up the basis of later analysis in three dimensions. These 2D systems have two independent symmetries: the $U(1)$ charge conservation and the mirror symmetry under the reflection $z\rightarrow -z$, where $z$ is normal to the 2D plane.  
Since this mirror symmetry is a $Z_2$ internal symmetry~\cite{internal}, 2D TCIs with mirror symmetry are synonymous to $U(1)\times Z_2$ SPT phases of fermions.  

In the absence of interactions, these 2D TCIs are classified by two integers $Z \oplus Z$, the Chern number $N$ and the mirror Chern number $n_M$ 
associated with occupied bands. Since the Chern number is defined without relying on the mirror symmetry, 
for our purpose it suffices to consider systems with $N=0$, for which the mirror Chern number $n_M$ is defined as the Chern number of the occupied bands 
with the mirror eigenvalue $ +1$~\cite{eval}.  For example, (001) thin films of SnTe and monolayers of IV-VI semiconductors are predicted to be 2D TCIs 
with $|n_M|=2$~\cite{liu,mn1,mn2,mn3}.


To study the classification of $U(1)\times Z_2$ SPT phases in the presence of interactions, we follow the general approach presented in the seminal work of  Lu and Vishwanath~\cite{lu} and analyze the stability of noninteracting edge states against interactions. The existence of edge states that can only be gapped by breaking the mirror symmetry signals a 2D SPT phase.  
To begin with, the low-energy Hamiltonian for edge states of noninteracting TCIs is given by   
\begin{equation}
H_{0} = \sum_av_{F} \int dx ( -i \psi_{a,R}^\dagger \partial_x \psi_{a,R} + i \psi_{a,L}^\dagger \partial_x \psi_{a,L} ). \label{h0}
\end{equation}
Here the fermion fields $\psi_{a,R/L}$ denote respectively the $a$-th right and left movers  ($a=1,...,n$), which transform differently under mirror:  
\begin{equation}
\label{eq:transform_psi}
M \psi_{a,R}^\dagger M^{-1} = \eta \psi_{a,R}^\dagger, \; 
M \psi_{a,L}^\dagger M^{-1} =  -\eta \psi_{a,L}^\dagger,  
\end{equation}
where $\eta= {\rm sgn}(n_M)$. 
The difference in mirror eigenvalues forbids single-particle backscattering between left and right movers; hence without interactions, gapless edge states 
are protected for any integer $n_M \neq 0$.  The velocity of different edge modes are chosen to be the same for simplicity; relaxing this condition will not affect any of our conclusions.   

We use bosonization to study the effect of interactions at the edge~\cite{CS1,CS2}. The bosonized Lagrangian for $H_0$ takes the form   
\begin{eqnarray}
L = \frac{1}{4\pi} K_{ij} \partial_x \phi_i \partial_t \phi_j
	-  \frac{1}{4\pi} v_F (\partial_x \phi_i)^2, \label{cs}
\end{eqnarray}
where $K$ is an integer-valued matrix given by  
\begin{equation}
K = 
\begin{pmatrix}
\bm{1}_{n \times n} & 0 \\
0 & -\bm{1}_{n \times n}
\end{pmatrix},
\end{equation} 
with $\bm{1}_{n \times n}$ being the $n \times n$ identity matrix. 
The boson field $\phi_i (x)$ satisfies the Kac-Moody algebra
\begin{equation}
[ \phi_i (x), \partial_{x'} \phi_j (x') ] 
= 2\pi i K^{-1}_{ij}  \delta(x-x'),  \label{phi}
\end{equation}
and the fermion fields $\psi^\dagger_{a,R/L}$ are given by
\beq
\psi^\dagger_{a,R} \sim e^{i\phi_{a}}, \; \psi^\dagger_{a,L} \sim e^{-i\phi_{n+a}}. \label{field}
\eeq

Electron interactions such as backscattering and umklapp processes can potentially gap the counter-propagating edge modes. 
These interaction terms are built from multi-electron creation and annihilation operators and are 
represented by cosine terms of the form $\cos ( \Phi_{\bm{L}}(x) + \alpha_{\bm{L}}(x))$, where  
the field $\Phi_{\bm{L}}(x) \equiv \bm{L}^T K \vec{\phi}(x)$ is defined by an integer-valued vector $\bm{L}$, and $\alpha_{\bm{L}}$ is an arbitrary phase. 
For our purpose, the interactions must preserve the charge conservation and mirror symmetry indispensable to 2D TCIs. 
It follows from eq.~\eqref{field} that charge conservation requires
\begin{equation}
\bm{L}^T \bm{t} = 0, \; \textrm{with } \bm{t}\equiv (\bm{1}_n, \bm{1}_n)^T. \label{u1}
\end{equation}
where $\bm{1}_n$ is the $n$-dimensional vector with all components equal to 1.  
For charge-conserving interactions, we further note the transformation law of the fermion field (\ref{eq:transform_psi}) under mirror symmetry 
implies
\begin{align}
M \Phi_{\bm{L}} M^{-1} 
= \Phi_{\bm{L}} +\eta  \frac{\pi}{2} \bm{L}^T \bm{m} , \; \textrm{with } \bm{m} \equiv (\bm{1}_n, -\bm{1}_n)^T.  
\end{align}
Hence  the condition for mirror symmetry requires
\begin{equation}
\label{eq:mirror}
\bm{L}^T \bm{m} \equiv 0  \mod 4. 
\end{equation}

To diagnose SPT phases, we consider sufficiently strong, symmetry-preserving interactions that completely gap the $2n$ edge modes. 
This can be achieved by adding to the edge Lagrangian (\ref{cs}) $n$ cosine terms~\cite{Levin1}: 
\beq
V = \sum_{a=1}^n \lambda_a  \cos ( \Phi_{\bm{L}_a}(x)), \label{cos}
\eeq 
where different fields $\Phi_{\bm{L}_a}$ are specified by a set of {\it linearly independent} integer-valued vectors $\bm{L}_a$, $a=1,...,n$.
To ensure that these fields can simultaneously have classical values, 
the commutator between any two of them must vanish. Since eq.~\eqref{phi} implies 
\begin{equation}
\label{eq:commutation}
[  \Phi_{\bm{L}_a}(x), \partial_{x'} \Phi_{\bm{L}_b}(x') ] 
= 2\pi i \bm{L}_a^T K \bm{L}_b   \delta (x-x'), 
\end{equation}
this commutativity condition requires 
\begin{equation}
\label{eq:orthogonal}
\bm{L}_a^T K \bm{L}_b = 0, 
\end{equation}
for any indices $a, b =1,...,n$. A set of such vectors $\{ \bm{L}_a \}$ will be referred to as a set of gapping vectors.  
As a general principle of bulk-boundary correspondence, the symmetry property of gapped edge states due to strong interactions reflects the topological property of the bulk. 
If the gapped edge preserves the $U(1) \times Z_2$ symmetry, the bulk is in a trivial phase, i.e., adiabatically connected to an atomic insulator.  

We now show this scenario occurs  
for edge states that have $n=4$ pairs of counter-propagating modes in the noninteracting limit. 
Such edge states can be gapped by interactions taking the bonsonized form eq.~\eqref{cos}, with the following set of gapping vectors $\bm{L}_a$:  
\begin{gather}
\bm{L}_1 = (1,1,0,0;-1,-1,0,0)^T, \nonumber \\
\bm{L}_2 = (0,0,1,1;0,0,-1,-1)^T, \nonumber \\
\bm{L}_3 = (1,-1,0,0;0,0,-1,1)^T, \nonumber \\
\bm{L}_4 = (1,0,1,0;-1,0,-1,0)^T.  \label{ln=4}
\end{gather}
It is easy to check that $\bm{L}_1,...,\bm{L}_4$ satisfy the symmetry conditions (\ref{u1}) and (\ref{eq:mirror}), as well as the commutativity condition (\ref{eq:orthogonal}). 
To motivate the choice of interactions (\ref{ln=4}), it is useful to regard four edge modes as two pairs of spinful Luttinger liquid in   
a two-leg fermion ladder system at half-filling. In the absence of inter-chain tunneling, the left- and right-moving modes   
have crystal momenta $\pm \pi/2$ and transform oppositely under the lattice translation: $c_R^\dagger \rightarrow i c_R^\dagger, c_L^\dagger \rightarrow - i c^\dagger_L$. 
This is identical to the mirror symmetry transformation property of TCI edge states (\ref{eq:transform_psi})---the only difference due to the factor $i$ can be eliminated by redefining the symmetry operator~\cite{eval}. Guided by this correspondence, we choose the interactions for $n=4$ edge states denoted by $\bm{L}_{1}$ and $\bm{L}_{2}$ to be the bosonized form of the Hubbard interaction in the two-leg ladder, and $\bm{L}_{3}$ and $\bm{L}_{4}$ to be the antiferromagnetic inter-chain coupling. 
The former opens up charge gap and effectively generates two spin chains; the latter opens up a spin gap and leads to a rung-singlet phase that is fully gapped and translationally invariant. 
Equivalently, the interactions (\ref{ln=4}) gap the $n=4$ edge states while preserving the mirror symmetry.  A detailed analysis can be found in the Supplementary Material~\cite{sm}.      
Therefore, we conclude that a noninteracting 2D TCI with mirror Chern number $n_M=\pm 4$ becomes trivial in the presence of interactions.  
The additive nature of SPT phases then implies the same conclusion holds for $n_M=4 k$, where $k$ is an integer.

Next we show case by case that the gapped edges of TCIs with $n=1$ and $2$ necessarily break the mirror symmetry spontaneously.   
First, $n=1$ edge states consist of  a pair of counter-propagating modes, which can be gapped by symmetry-allowed Umklapp interactions that backscatter  
an even number of electrons from left to right movers, described by $\cos (2 k \Phi_{\bm{L}})$ with $\bm{L}=(1,-1)^T$. 
The gap generation then implies $\Phi_{\bm{L}}$ is pinned, i.e., $\langle e^{i \Phi_{\bm{L}}}\rangle \neq 0$. This signals spontaneous mirror symmetry breaking, as can be seen from (\ref{eq:mirror}).


For $n=2$,  by an exhaustive enumeration, we find two types of symmetry-preserving two-body interactions that gap the edge states, which are specified 
by two sets of gapping vectors $\{ \bm{L}_1, \bm{L}_2\}$ and $\{ \bm{L}_1, \tilde{\bm{L}}_2\}$ respectively, with   
$
\bm{L}_1 = (1, 1; -1, -1)^T, 
\bm{L}_2 = (1, -1; -1, 1)^T
$
and 
$
\tilde{\bm{L}}_2 = (1, -1; 1, -1)^T. 
$
We further note that the second type of interaction becomes equivalent to the first after 
redefining the flavor index of the left-movers $\psi^\dagger_{1L} \leftrightarrow \psi^\dagger_{2L}$. 
Hence only the first type of interaction needs to be considered. In terms of the electron operators, this interaction takes the form
\begin{align}
V  =& \lambda_1 ( \psi_{1R}^\dagger \psi_{2R}^\dagger \psi_{2L} \psi_{1L} + \text{h.c.}) \nonumber \\
& + \lambda_2  ( \psi_{1R}^\dagger \psi_{2L}^\dagger \psi_{1L} \psi_{2R} + \text{h.c.}). \label{v12}
\end{align}
Both terms conserve the number of fermions in each flavor (denoted by $a=1,2$) and commute with each other. 
The first term is an Umklapp process that backscatters two electrons with different flavors, 
and  the second term flips the flavor of left and right movers simultaneously.
It is convenient to introduce boson fields for each flavor: 
$
\varphi_a= (\phi_{a,R} + \phi_{a,L})/2
$
and
$
\theta_a = (\phi_{a,R}  - \phi_{a, L})/2 
$, with $n_a = \partial_x \theta_a$ being the density of electrons in flavor $a$. 
Equation \eqref{v12} then becomes 
\begin{gather}
V = \lambda_1 \cos(2\theta_1 + 2\theta_2) +  \lambda_2 \cos(2\theta_1 - 2\theta_2).
\end{gather}
In the presence of this interaction, the edge becomes gapped when the fields $\theta_1$ and $\theta_2$ are both pinned.  
This leads to nonzero expectation values of single-particle backscattering operators: 
$\langle e^{i 2\theta_1} \rangle \sim \langle \psi^\dagger_{1R} \psi_{1L} \rangle \neq 0$ and 
$\langle e^{i 2\theta_2} \rangle  \sim \langle \psi^\dagger_{2R} \psi_{2L} \rangle \neq 0$, which implies spontaneous mirror symmetry breaking.

The above edge state analysis shows that noninteracting TCIs with mirror Chern number $n_M=\pm 1$ and $\pm 2$ remain topologically nontrivial in the presence of interactions, 
contrary to the previous case of $n_M=4k$. Therefore, we conclude that interactions reduce the classification 
of 2D TCIs protected by mirror symmetry, or $U(1)\times Z_2$ SPT phases, from $Z$ to $Z_4$. 

In addition to  its theoretical value, the above result has important implications for thin films/monolayers of SnTe and other IV-VI semiconductors, which are predicted to be 2D TCIs with 
$|n_M|=2$ by band structure calculations~\cite{liu,mn1,mn2,mn3}. Our analysis shows that interactions of the form (\ref{v12}) can qualitatively change the properties of $n=2$ edge states.    
At generic filling, only the flavor-flipping $\lambda_2$ term is allowed by momentum conservation and it is relevant for repulsive Luttinger interaction from the renormalization group analysis~\cite{sm}.  
As a result, there appears a gap in the flavor sector while the charge sector remains gapless and fluctuates.  
Boundaries and impurities affect the charge mode by pinning a fluctuating charge density wave, which can be detected by STM measurement similar to the case of Luther-Emery liquid with a spin gap~\cite{CDW}.

{\bf Interacting TCIs in three dimensions}: We now turn to TCIs in three dimensions, protected by a single mirror symmetry, say $x \rightarrow -x$.  
Within band theory, one can define the mirror Chern number $n_M$ on the 2D plane $k_x=0$ in $\bm{k}$-space, which is invariant under this reflection. 
The integer $n_M$ thus classifies 3D noninteracting TCIs~\cite{hsieh, chiu, furusaki, sato}. The hallmark surface states, present on crystal surfaces symmetric under mirror, consist of $n=|n_M|$ Dirac cones:
\begin{equation}
H_0 = \sum_{a=1}^n v_F \int d \bm{r} \;  \psi^\dagger_a(\bm{r}) (- i \partial_x s_y + i \partial_y s_x ) \psi_a(\bm{r}),
\end{equation}
where $\psi^\dagger_a = (\psi^\dagger_{a\uparrow}, \psi^\dagger_{a \downarrow})$ is a two-component fermion field. 
Reflection acts on both electron's coordinate and spin as follows:
\begin{eqnarray}
M \psi^\dagger_a (x, y) M^{-1} =  s_x \psi^\dagger_a(-x, y).  
\end{eqnarray}
The mirror symmetry forbids any Dirac mass term $\psi^\dagger_a s_z \psi_b$, and thus protects these $n$ flavors of gapless Dirac fermions.   

Can the above Dirac fermion surface states be gapped by interactions without breaking the charge conservation and mirror symmetry? 
Finding the answer to this questions will hold the key to the classification of interacting TCIs in three dimensions. This is a challenging task 
requiring non-perturbative approach to strongly interacting Dirac fermions in two dimensions.  


We now demonstrate explicitly that interactions can turn surface states with $n=8$ flavors of 
Dirac fermions into a gapped and mirror symmetric phase without intrinsic topological order (i.e., without fractional excitations). 
Such a completely trivial surface phase is constructed as follows. First, we introduce a spatially alternating Dirac mass term to $H_{0}$: 
\begin{align}
H_m = \int d \bm{r} \; m (x)  \left(  \sum_{a=1}^4 \psi_a^\dagger(\bm{r}) s_z \psi_a(\bm{r}) -  \sum_{a=5}^8 \psi_a^\dagger(\bm{r}) s_z \psi_a(\bm{r}) \right) , \label{dm1}
\end{align}
where $m(x)$ is a periodic function of $x$ that alternates between $m_0$ and $-m_0$,  
\begin{equation}
m(x)  =  
\begin{cases}
m_0  &  \text{for } (2k-1) L <x < 2k L , \\
- m_0 & \textrm{for }  2kL <x < (2k+1)L. 
\end{cases} \label{dm2}
\end{equation}
Importantly, the resulting periodic array of Dirac mass domains preserves the mirror symmetry, because $m(x)=-m(-x)$ and $M \psi_a^\dagger(x,y) s_z \psi_a(x,y) M^{-1} =  -\psi_a^\dagger(-x,y) s_z \psi_a(-x, y)$. 

\begin{figure}
\centering
\includegraphics[width=0.8\hsize]{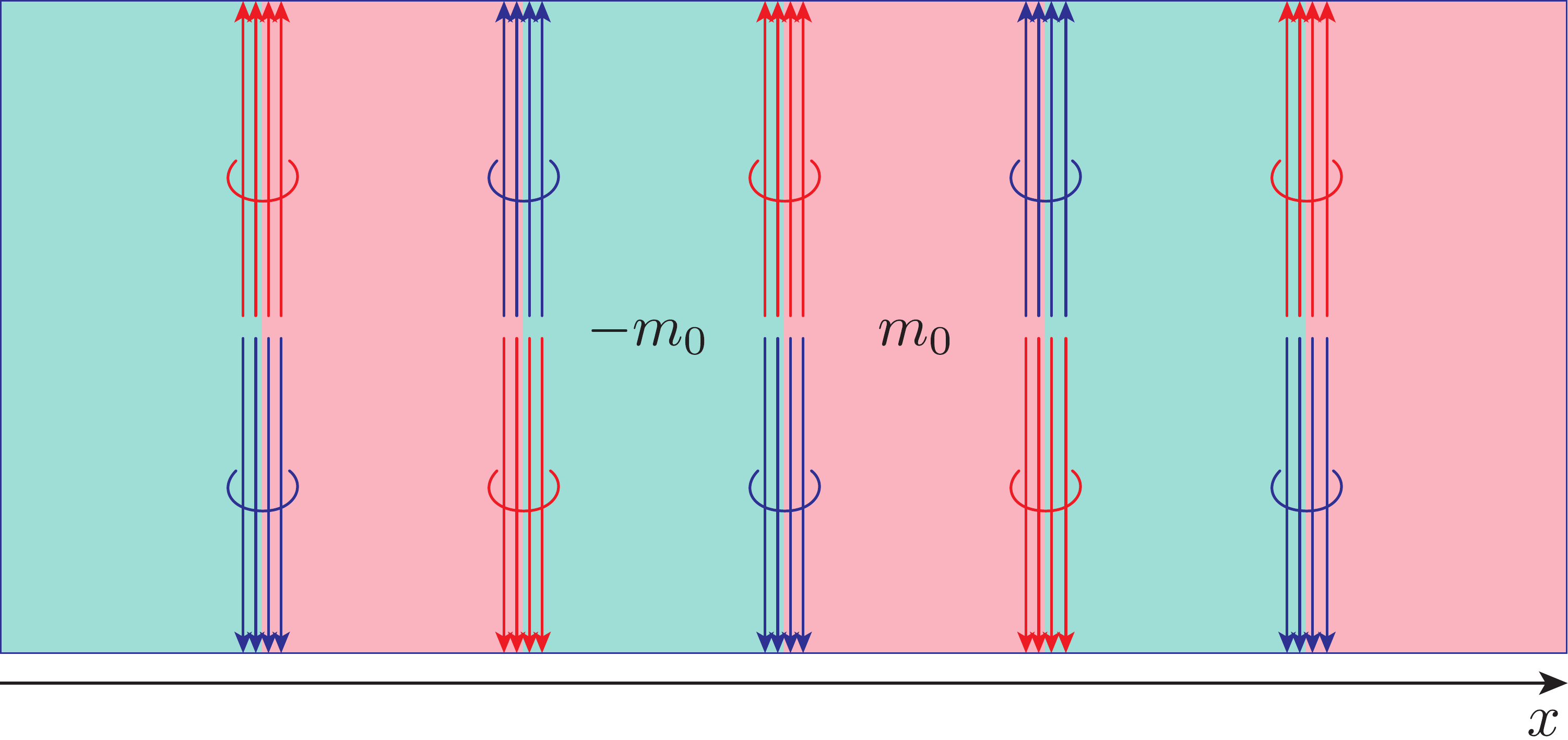}
\caption{(Color online) 
Periodic array of 1D domain wall fermions, generated by spatially alternating Dirac masses to 8 flavors of 2D Dirac fermions, see Eq.(\ref{dm1},\ref{dm2}). 1D chiral fermion modes in flavors $1,...,4$ (red arrows) and flavors $5,...,8$ (blue arrows) propagate in opposite directions along a domain wall. Counter-propagating chiral fermions have opposite mirror eigenvalues $\pm 1$. Each domain wall becomes gapped under the interaction (\ref{cos},\ref{ln=4}), thus leading to a gapped and mirror-symmetric 2D phase.  
}
\label{array}
\end{figure}

When the Dirac mass $m_0$ is large and the width of the domain $L$ is large, the low-energy degrees of freedom are confined to the domain walls at $x=kL$, where 
the Dirac mass changes sign.  As is well-known, the mass domain wall of a 2D Dirac fermion hosts a 1D chiral fermion mode, whose directionality is reversed upon 
changing the signs of the Dirac masses on both sides. Therefore, our setup  described by $H_0 + H_m$ hosts an array of 1D domain wall fermions, one per flavor.  
On each domain wall, chiral fermions in flavors $1,...,4$ and those in flavors $5,...,8$ move in opposite directions, and importantly, have opposite mirror eigenvalues $\pm 1$ 
under the spatial reflection interchanging the two sides of the domain wall, as shown in Fig.~\ref{array}. 
 
 We now draw a connection between the domain wall states on the surface of 3D TCIs to the edge states of 2D TCIs: both are 1D system of counter-propagating fermions with opposite mirror eigenvalues. Without interactions, the locking between the directionality and mirror eigenvalue forbids single-particle backscattering, leaving such 1D system gapless. 
However, as we have shown earlier, the interaction given by eqs.~\eqref{cos} and  \eqref{ln=4} opens up a gap when there are four pairs of counter-propagating modes. 
Applying this interaction to each domain wall that we set up on the surface of noninteracting TCIs then gaps the entire surface state states with $n=8$ Dirac fermions, while preserving the mirror symmetry $x \rightarrow -x$.       
We have thus explicitly constructed, using a periodic array of domain walls, a completely trivial and gapped surface, the existence of which then implies noninteracting TCIs with mirror Chern number $n_M=8k$ become trivial in the presence of interactions.  

\begin{figure}
\centering
\includegraphics[width=\hsize]{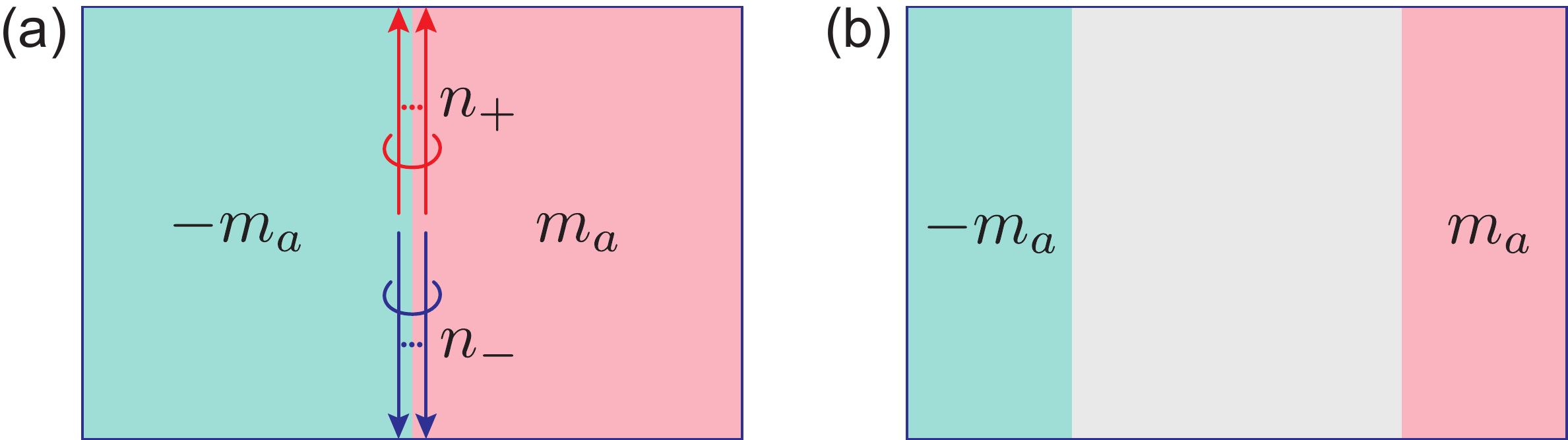}
\caption{(Color online) 
(a) A mass domain wall setup on a 3D TCI surface consisting of  $n$ flavors of Dirac fermions. One-dimensional chiral fermions reside at the domain wall at $x=0$, with $n_+$ ($n_- = n -n_-$) modes move in the $+y$ ($-y$) direction, depending on the signs of Dirac masses $m_1, ..., m_n$. 
(b) The domain wall in (a) is expanded to a wide region, sandwiched between semi-infinite regions on the left and on the right, with opposite Dirac masses.    
Importantly, (a) and (b) are both symmetric under mirror $x \rightarrow -x$ and topologically equivalent.  
For $n \neq 8k$, the domain wall in (a), hence  the middle region in (b) as well, cannot be gapped and mirror-symmetric. 
}
\label{setup}
\end{figure}

Next, let us consider surface states of TCIs with $n_M \neq 8k$.  Below we prove by contradiction that interactions cannot generate a gapped, mirror symmetric and non-fractionalized phase for these surface states~\cite{disorder}. 
Suppose such a trivial gapped phase exists,  it must be adiabatically connectable to a massive Dirac fermion phase, where the Dirac masses are generated by external 
mirror symmetry breaking perturbations. This motivates us to consider a sandwich setup shown in Fig.~\ref{setup}b, 
where this trivial phase takes up the region $|x| < L$; to its right is a massive phase with a set of Dirac masses $\{ m_a \}$; and to its left the mirror image, a massive phase with opposite Dirac masses $\{-m_a \}$. By construction, this sandwich setup is symmetric under the reflection $x \rightarrow -x$.

We choose $L$ to be much larger than the correlation length of the trivial gapped phase and let the surface Hamiltonian vary slowly with $x$ across the interface at $x=\pm L$, so that 
the trivial gapped phase (presumed to exit) adiabatically evolves into the massive Dirac fermion phase, without closing gap at the interface.  
Therefore, the surface is everywhere gapped and as a whole preserves the mirror symmetry. 

On the other hand, the sandwich setup is topologically equivalent to a domain wall between two domains with opposite Dirac masses (Fig.~\ref{setup}a). 
Without interactions, this domain wall hosts $n= |n_M|$ flavors of 1D chiral fermions, with $n_+$ flavors and $n_-$ flavors moving in opposite directions and carrying opposite mirror eigenvalues.  Here $n_+$ ($n_-$) is the number of Dirac fermions with $m_a>0$ ($m_a<0$), and $n_+ + n_- =n$. Importantly, for $n_+\neq n_-$, the domain wall must be gapless 
due to the presence of a net chirality, and for $n_+ = n_- =n /2 \neq 4k$, we have shown earlier that the domain wall cannot be trivially gapped by interactions either.   
This result of the domain wall contradicts with that of the sandwich setup, which is deduced to be gapped under the assumption that a trivial gapped surface is allowed on $n \neq 8k$ TCI surfaces.  
This contradiction proves the assumption wrong. Instead, 3D TCIs with mirror Chern number $n_M \neq 8k$ cannot have a trivial gapped surface and hence remain topologically nontrivial in the presence of interactions. Putting everything together, we conclude that interactions reduce the classification of 3D TCIs with mirror symmetry from $Z$ to $Z_8$.   
  
In addition to reducing the classification of noninteracting TCIs,  interactions may also enable new TCI phases that do not exist in free fermion systems, as recently found in other symmetry classes~\cite{wang2,hughes}. We leave this interesting problem of interaction-enabled TCIs with mirror symmetry for future study.

\begin{acknowledgments}
We thank Senthil Todadri and Chong Wang for interesting discussions. HI is supported by the ALPS program, Grant-in-Aid for JSPS Fellows, and LF is supported by David and Lucile Packard Foundation.  
\end{acknowledgments}

%
%

\begin{center}
{\large\bf Supplemental Material}
\end{center}


In the Supplemental Material, we provide the approach to gapping vectors for $n=4$ case, and the renormalization group (RG) analysis for $n=2$ case. 
We utilize the Tomonaga-Luttinger liquid description in the following analyses, which we introduce first. 

In the Tomonaga-Luttinger theory, the fermion field $\psi^\dagger_{a,R/L}$ is given by~\cite{bosonization}
\begin{gather}
\psi^\dagger_{a,R} = \frac{1}{\sqrt{2\pi\alpha}} e^{i\phi_{a}} = \frac{1}{\sqrt{2\pi\alpha}} e^{i(\varphi_a + \theta_a)}, \notag \\
\psi^\dagger_{a,L} = \frac{1}{\sqrt{2\pi\alpha}} e^{-i\phi_{a+n}} = \frac{1}{\sqrt{2\pi\alpha}} e^{i(\varphi_a - \theta_a)}.
\end{gather}
$\alpha$ is an infinitesimal convergence factor. We neglect the Klein factor here. 
The two fields $\varphi_a(x)$ and $\theta_a(x)$ satisfy the commutation relation
\begin{equation}
[ \theta_a(x), \varphi_a(x) ] = i\frac{\pi}{2} \text{sgn}(x-x'),
\end{equation}
and transform under the mirror symmetry as 
\begin{equation}
\label{eq:transform_luttinger}
M \varphi_a(x) M^{-1} = \varphi_a(x), \;
M \theta_a(x) M^{-1} = \theta_a(x)+\frac{\pi}{2}. 
\end{equation}
The electron density for the right and left movers are given by $n_{a,R/L} = \partial_x \phi_{a(+n)}/(2\pi)$ and thus the total electron density of $a$-th pair is $n_a = n_{a,R} + n_{a,L} = \partial_x \theta_a/\pi$. 
With the definition above, the bosonized Hamiltonian without gap-opening scatterings is
\begin{equation}
H^0 = \sum_{a=1}^n H_a^0 (v_a, K_a)
\end{equation}
with
\begin{align}
H_a^0 (v_a, K_a) = \frac{v_a}{2\pi} \int dx \left[ K_a (\partial_x \varphi_a)^2 + \frac{1}{K_a} (\partial_x \theta_a)^2 \right]. 
\end{align}
The forward scattering terms, $g_2$ and $g_4$, are included through the Luttinger parameter $K_a$ and renormalized velocity $v_a$, defined as 
\begin{gather}
K_a = \sqrt{\frac{1+(g_{4,a}-g_{2,a})/(2\pi v_{F,a})}{1+(g_{4,a}+g_{2,a})/(2\pi v_{F,a})}}, \\
v_a = v_{F,a} \sqrt{\left( 1+\frac{g_{4,a}}{2\pi v_{F,a}} \right)^2 - \left( \frac{g_{2,a}}{2\pi v_{F,a}} \right)^2}.
\end{gather}

\subsection*{Gapped states for $n=4$: rung-singlet phase}

In this section, we elaborate on the interaction Hamiltonian for $n=4$ in the main text, which gaps four pairs of counter-propagating edge modes. 
It is instructive to make an analogy between such edge states and the low-energy states of a two-leg fermion ladder system at half-filling. 
In the noninteracting limit, each chain is described by the tight-binding Hamiltoinan
\begin{align}
H_0 =& -t \sum_{j,\sigma} (c^\dagger_{j+1,\sigma} c_{j,\sigma} + \text{H.c.}),   
\end{align}
where $j$ is a site index and the spin $\sigma = \uparrow, \downarrow$. 
Each chain supports spin-degenerate left- and right-moving modes, which have crystal momenta $\pm \pi/2$ respectively and transform oppositely under the lattice translation: 
\begin{equation}
c_R^\dagger \rightarrow i c_R^\dagger, c_L^\dagger \rightarrow - i c^\dagger_L.
\end{equation} 
This is equivalent to the mirror symmetry transformation property of TCI edge states stated in the main text. 

We now add on-site Hubbard interaction to each chain, 
\begin{align}
H_U=  U \sum_j \left( n_{j,\uparrow} - \frac{1}{2} \right) \left( n_{j,\downarrow} - \frac{1}{2} \right). 
\end{align}
For $U>0$, the repulsive interaction opens up a charge gap. As  a result, at low energy each chain is equivalent to a spin-1/2 chain, as in the antiferromagnetic Heisenberg spin model. 
By further introducing antiferromagnetic inter-chain coupling, 
we obtain the rung-singlet phase of two coupled spin chains, which is gapped and translationally invariant.  Back to our original problem, this corresponds a gapped edge preserving the mirror symmetry. 

The remaining task is to derive the bosonized form of the microscopic Hubbard and spin interactions in terms of the left- and right-moving fermion fields. The bosonized form of the Hubbard model is
\begin{equation}
H = H_\rho + H_\sigma, 
\end{equation}
where the ``charge'' $(\rho)$ and ``spin'' $(\sigma)$ degrees of freedom are separated to give 
\begin{align}
\label{um}
H_\rho = H^0_\rho (v_\rho, K_\rho) - \frac{2U}{(2\pi\alpha)^2} \int dx \cos(2\sqrt{2}\theta_\rho), \\
\label{bs}
H_\sigma = H^0_\sigma (v_\sigma, K_\sigma) + \frac{2U}{(2\pi\alpha)^2} \int dx \cos(2\sqrt{2}\theta_\sigma). 
\end{align}
The fields are defined by
\begin{equation}
\theta_{\rho/\sigma} = \frac{1}{\sqrt{2}} (\theta_1 \pm \theta_2), \; \varphi_{\rho/\sigma} = \frac{1}{\sqrt{2}} (\varphi_1 \pm \varphi_2),
\end{equation}
and the renormalized velocities and the Luttinger parameters are 
\begin{align}
v_{\rho/\sigma} = v_F \left( 1 \pm \frac{U}{\pi v_F} \right)^{1/2}, \; K_{\rho/\sigma} = K \left( 1 \pm \frac{U}{\pi v_F} \right)^{-1/2}. 
\end{align}
The Umklapp process in eq.~\eqref{um} is relevant for $K_\rho < 1$ at half-filling, and the backscattering in eq.~\eqref{bs} is relevant for $K_\sigma < 1$ at generic filling. 
For $U>0$ and half-filling, the Umklapp process is relevant and opens the charge gap. 
This Umklapp interaction corresponds to two gapping vectors:
\begin{gather}
\bm{L}_1 = (1,1,0,0;-1,-1,0,0)^T, \notag \\
\bm{L}_2 = (0,0,1,1;0,0-1,-1)^T.
\end{gather}

Still the ``spin'' modes remain gapless. 
Now we can regard the two gapless modes as a two-leg ladder of spin chains~\cite{millis}. 
The Jordan-Wigner transformation for a  spin chain $i$ $(i=1,2)$ and subsequent bosonization give
\begin{gather}
S^z_i (x) = \frac{1}{\pi} \partial_x \theta_i(x) + \frac{(-1)^x}{\pi\alpha} \cos(2\theta_i(x)), \notag \\
S^+_i (x) = \frac{e^{i\varphi_i(x)}}{\sqrt{2\pi\alpha}} [ (-1)^x + \cos(2\theta_i(x)) ].
\end{gather}
($x$ is actually defined on a lattice $x=aj$ and finally we take the continuum limit. Here we take the lattice constant $a=1$ for simplicity.) 
We assume the spin chains are written as the Heisenberg model 
\begin{equation}
H_i = J \sum_j \bm{S}_{i,j} \cdot \bm{S}_{i,j+1}, 
\end{equation}
with $i$ is the spin chain index and $j$ denotes a site, i.e., a rung. The interchain coupling acts in a rung as 
\begin{equation}
H_\perp = J_\perp^{xy} \sum_j (S_{1,j}^x S_{2,j}^x + S_{1,j}^y S_{2,j}^y) + J_\perp^z \sum_j S_{1,j}^z S_{2,j}^z. 
\end{equation}
The total Hamiltonian of the two-leg spin ladder is 
\begin{equation}
H = H_1 + H_2 + H_\perp.
\end{equation}
It can be decomposed as 
\begin{equation}
H = H_s + H_a, 
\end{equation}
where
\begin{align}
\label{sym}
H_s &= H^0_s (u_s,K_s) + \frac{2J_\perp^z}{(2\pi\alpha)^2} \int dx \cos (2\sqrt{2}\theta_s), \\
\label{asym}
H_a &= H^0_a (u_a, K_a) + \frac{2J_\perp^z}{(2\pi\alpha)^2} \int dx \cos (2\sqrt{2}\theta_a) \notag \\
&\quad + \frac{2\pi J_\perp^{xy}}{(2\pi\alpha)^2} \int dx \cos (\sqrt{2}\varphi_a).
\end{align}
$s$ denotes the symmetric part and $a$ the antisymmetric part, defined by 
\begin{equation}
\theta_{s/a} = \frac{1}{\sqrt{2}} (\theta_1 \pm \theta_2), \; \varphi_{s/a} = \frac{1}{\sqrt{2}} (\varphi_1 \pm \varphi_2). 
\end{equation}
The renormalized velocity and Luttinger parameters are 
\begin{equation}
v_{s/a} = J \left( 1 \pm \frac{K J_\perp^z}{2\pi J} \right), \; K_{s/a} = K \left( 1 \mp \frac{K J_\perp^z}{2\pi J} \right). 
\end{equation}
Here $K=1/2$ since we assume the Heisenberg model for each spin chains. 
For the symmetric part \eqref{sym}, $\cos (2\sqrt{2}\theta_s)$ is relevant for $K_s <1$ due to the RG analysis. 
In contrast, for the antisymmetric part \eqref{asym}, two cosine terms compete but from the RG analysis $\cos (2\sqrt{2}\theta_a)$ is relevant for $K_a < 1/2$ and $\cos (\sqrt{2}\varphi_a)$ for $K_a >1/2$. 
If we assume the antiferromagnetic interchain coupling, i.e., $J_\perp^z > 0$, the Luttinger parameter becomes $K_a > 1/2$ and thus $\cos (\sqrt{2}\varphi_a)$ is relevant. 
Now the two fields are pinned, the system becomes completely gapped. 

Since we assume that the interchain coupling is antiferromagnetic, two spins in a rung form a singlet; it is called a \textit{rung-singlet phase}. 
The way of gapping two-leg ladders is related to the Haldane gap for integer spin chains. 

The next step is to determine the corresponding gapping vectors. 
Note that the fields in the two-leg spin ladder model come from the ``spin'' modes of the Hubbard model we considered first. 
Thus we should replace 
\begin{gather}
\theta_1 \to \theta_1 - \theta_2, \; \theta_2 \to \theta_3 - \theta_4, \notag \\
\varphi_1 \to \varphi_1 - \varphi_2, \; \varphi_2 \to \varphi_3 - \varphi_4, 
\end{gather} 
to obtain $\bm{L}$. 
By these replacement, we obtain the gapping vectors
\begin{gather}
\bm{L}_3 = (1,-1,1,-1;-1,1,-1,1)^T, \notag \\
\bm{L}_4 = (1,-1,-1,1;1,-1,-1,1)^T, 
\end{gather}
where $\bm{L}_3$ and $\bm{L}_4$ correspond to $\cos (2\sqrt{2}\theta_s)$ and $\cos (\sqrt{2}\varphi_a)$, respectively. 

%

Now we have four gapping vectors $\bm{L}_a$ $(a=1,...,4)$. However, we should confirm the absence of spontaneous symmetry breaking. 
An SPT phase and a trivial phase (such as an atomic insulator)  are distinguished by the symmetry property of the gapped edge states.  
While a trivial  phase permits a gapped and symmetry-preserving edge,  
edge states of a SPT phase, if  gapped, must spontaneously break the protecting symmetry. 
As shown by Levin and Stern~\cite{Levin2}, spontaneous symmetry breaking may (but not necessarily) occur 
when a linear combination of gapping vectors $\sum_i c_i \bm{L}_i$ for the coefficients $\{ c_i \}$ with no common divisors is nonprimitive, i.e., 
\begin{equation}
\label{eq:primitive}
\sum_i c_i \bm{L}_i = c \bm{L}
\end{equation}
and the integer $c$ is larger than $1$. 
In this case, the set of pinned fields $\{ \Phi_{\bm{L}_i} \}$, which themselves are symmetry-preserving, also freezes the field $\Phi_{\bm{L}}$.  
The latter may or may not break the original symmetry of the system, which needs to be checked case by case. 
Conversely,  if $\sum_i c_i \bm{L}_i$ is primitive for any coefficients with no common divisors, spontaneous symmetry breaking is guaranteed to be absent. 

The set of four gapping vectors $\bm{L}_a$ $(a=1,...,4)$ is not primitive. 
Thus we define a new primitive set $\bm{L}'_a$ from the linear combinations of $\bm{L}_a$: 
\begin{gather}
\bm{L}'_1 = \bm{L}_1 = (1,1,0,0;-1,-1,0,0)^T, \notag \\
\bm{L}'_2 = \bm{L}_2 = (0,0,1,1;0,0,-1,-1)^T, \notag \\
\bm{L}'_3 = \frac{1}{2} ( \bm{L}_3 + \bm{L}_4 ) = (1,-1,0,0;0,0,-1,1)^T, \notag \\
\bm{L}'_4 = \frac{1}{2} ( \bm{L}_1 + \bm{L}_2 + \bm{L}_3 ) = (1,0,1,0;-1,0,-1,0)^T. 
\end{gather}
The new set $\bm{L}'_a$ respects the $U(1) \times Z_2$ symmetry, and the primitivity ensures the absence of spontaneous symmetry breaking.  
Therefore we conclude that the edge modes can be gapped out without breaking symmetry for $n=4$. 
The set of gapping vectors $\bm{L}'_a$ is given in eq.~(13) in the main text.


\subsection*{RG analysis for $n=2$ edge modes}

As mentioned in the main text, we find only two sets of symmetry-preserving gapping vectors for $n=2$: 
\begin{gather}
\bm{L}_1 = (1, 1; -1, -1)^T, \notag \\
\bm{L}_2 = (1, -1; -1, 1)^T, 
\end{gather}
and 
\begin{gather}
\bm{L}_1 = (1, 1; -1, -1)^T, \notag \\
\tilde{\bm{L}}_2 = (1, -1; 1, -1)^T.
\end{gather}
In the following, we will consider the two sets in the Tomonaga-Luttinger description, and will analyze their relevance by RG calculation. 
At the beginning, we assume the two equivalent edge modes by setting $v_1 = v_2 = v$ and $K_1 = K_2 = K$. 
When two pairs of edge modes exist, two types of forward scatterings connecting two copies are allowed: 
\begin{gather}
V'_2 = g'_2 \int dx ( \psi^\dagger_{1R} \psi_{1R} \psi^\dagger_{2L} \psi_{2L}
	+ \psi^\dagger_{1L} \psi_{1L} \psi^\dagger_{2R} \psi_{2R}), \\
V'_4 = g'_4 \int dx ( \psi^\dagger_{1R} \psi_{1R} \psi^\dagger_{2R} \psi_{2R}
	+ \psi^\dagger_{1L} \psi_{1L} \psi^\dagger_{2L} \psi_{2L}).
\end{gather}
Bosonizing the two processes $V'_2$ and $V'_4$, we obtain 
\begin{equation}
H = \frac{1}{2\pi} \int dx [ (\partial_x \vec{\varphi})^T M_\varphi (\partial_x \vec{\varphi}) + (\partial_x \vec{\theta})^T M_\theta (\partial_x \vec{\theta}) ],
\end{equation}
where $\vec{\varphi} = (\varphi_1, \varphi_2)^T$, $\vec{\theta} = (\theta_1, \theta_2)^T$, and the matrices $M_\varphi$ and $M_\theta$ are given by
\begin{gather}
M_\varphi = 
\begin{pmatrix}
vK & (g'_4-g'_2)/2\pi \\ (g'_4-g'_2)/2\pi & vK
\end{pmatrix}, \\
M_\theta = 
\begin{pmatrix}
v/K & (g'_4+g'_2)/2\pi \\ (g'_4+g'_2)/2\pi & v/K
\end{pmatrix}.
\end{gather}
The matrices $M_\varphi$ and $M_\theta$ can be diagonalized simultaneously to obtain 
\begin{align}
H =& \frac{v_+}{2\pi} \int dx \left[ K_+ (\partial_x \varphi_+)^2 + \frac{1}{K_+} (\partial_x \theta_+)^2 \right] \notag \\
&+ \frac{v_-}{2\pi} \int dx \left[ K_- (\partial_x \varphi_-)^2 + \frac{1}{K_-} (\partial_x \theta_-)^2 \right]
\end{align}
with the new Luttinger parameter
\begin{equation}
K_\pm = \sqrt{\frac{vK \pm (g'_4-g'_2)/2\pi}{v/K \pm (g'_4+g'_2)/2\pi}},
\end{equation}
and the renormalized velocity
\begin{equation}
v_\pm = \sqrt{\left( vK \pm \frac{g'_4-g'_2}{2\pi} \right) \left( \frac{v}{K} \pm \frac{g'_4+g'_2}{2\pi} \right)}.
\end{equation}
The fields $\varphi_\pm$ and $\theta_\pm$ are defined by 
\begin{equation}
\varphi_\pm = \frac{1}{\sqrt{2}} (\varphi_1 \pm \varphi_2), \quad 
\theta_\pm = \frac{1}{\sqrt{2}} (\theta_1 \pm \theta_2).
\end{equation}

First we consider the scattering processes denoted by $\bm{L}_1$ and $\bm{L}_2$. 
The two scattering processes are written as 
\begin{gather}
V_1 = g_u \int dx ( e^{-4ik_F x} \psi_{1R}^\dagger \psi_{2R}^\dagger \psi_{2L} \psi_{1L} + \text{h.c.} ), \\
V_2 = g_b \int dx ( \psi_{1R}^\dagger \psi_{2L}^\dagger \psi_{1L} \psi_{2R} + \text{h.c.} ).
\end{gather}
$V_1$ is an Umklapp process occurring at half-filling $k_F = \pi/2$ and $V_2$ is a backscattering allowed at generic filling. 
Their bosonized forms are 
\begin{gather}
V_1 = \frac{2g_u}{(2\pi\alpha)^2} \int dx \cos(2\theta_1 + 2\theta_2), \\
V_2 = \frac{2g_b}{(2\pi\alpha)^2} \int dx \cos(2\theta_1 - 2\theta_2),
\end{gather}
or by using $\theta_\pm$
\begin{gather}
V_1 = \frac{2g_u}{(2\pi\alpha)^2} \int dx \cos(2\sqrt{2}\theta_+), \\
V_2 = \frac{2g_b}{(2\pi\alpha)^2} \int dx \cos(2\sqrt{2}\theta_-).
\end{gather}
The RG analysis shows that $V_1$ is relevant for $K_+<1$ and $V_2$ for $K_-<1$. 
When a scattering process is relevant, it pins the field $\theta_\pm$ and generates a gap. 
The pinning of $\theta_\pm$ leads to the mass $\Delta_\pm$, estimated as $\Delta_+ \approx (v_+/\alpha)(g_u)^{1/(2-2K_+)}$ and $\Delta_- \approx (v_-/\alpha)(g_b)^{1/(2-2K_-)}$. 
This situation resembles the charge-spin separation of conventional spinful 1D systems. 
The fields $\varphi_+$ and $\theta_+$ correspond to the charge degrees, and $\varphi_-$ and $\theta_-$ to the spin degrees. 
The charge sector is gapped by the Umklapp process and the spin sector by the backscattering process.

The mirror symmetry restricts the simultaneous gap opening of $\Delta_+$ and $\Delta_-$ because the pinning of $\theta_\pm$ means the pinning of $\theta_{1,2}$. 
Since $\theta_1$ and $\theta_2$ have a periodicity of $\pi$, $\theta_1+\theta_2$ is pinned at either 0 or $\pi$ (mod $2\pi$) for $g_u<0$, and either $\pi/2$ or $3\pi/2$ (mod $2\pi$) for $g_u>0$. 
Similar consideration applies for the backscattering process, which pins $\theta_1-\theta_2$ and its value depends on the sign of $g_b$. 
Therefore $\theta_{1,2}$ have expectation values of either 0, $\pi/4$, $\pi/2$, or $3\pi/4$, depending on the signs of $g_u$ and $g_b$, and hence the mirror symmetry is spontaneously broken. 
It is also leads to non-zero expectation values of single-particle backscattering $\langle e^{i2\theta_1} \rangle \sim \langle \psi_{1R}^\dagger \psi_{1L} \rangle \neq 0$ and $\langle e^{i2\theta_2} \rangle \sim \langle \psi_{2R}^\dagger \psi_{2L} \rangle \neq 0$, which is prohibited by the mirror symmetry.

\begin{figure}[t]
\centering
\includegraphics[width=0.8\hsize]{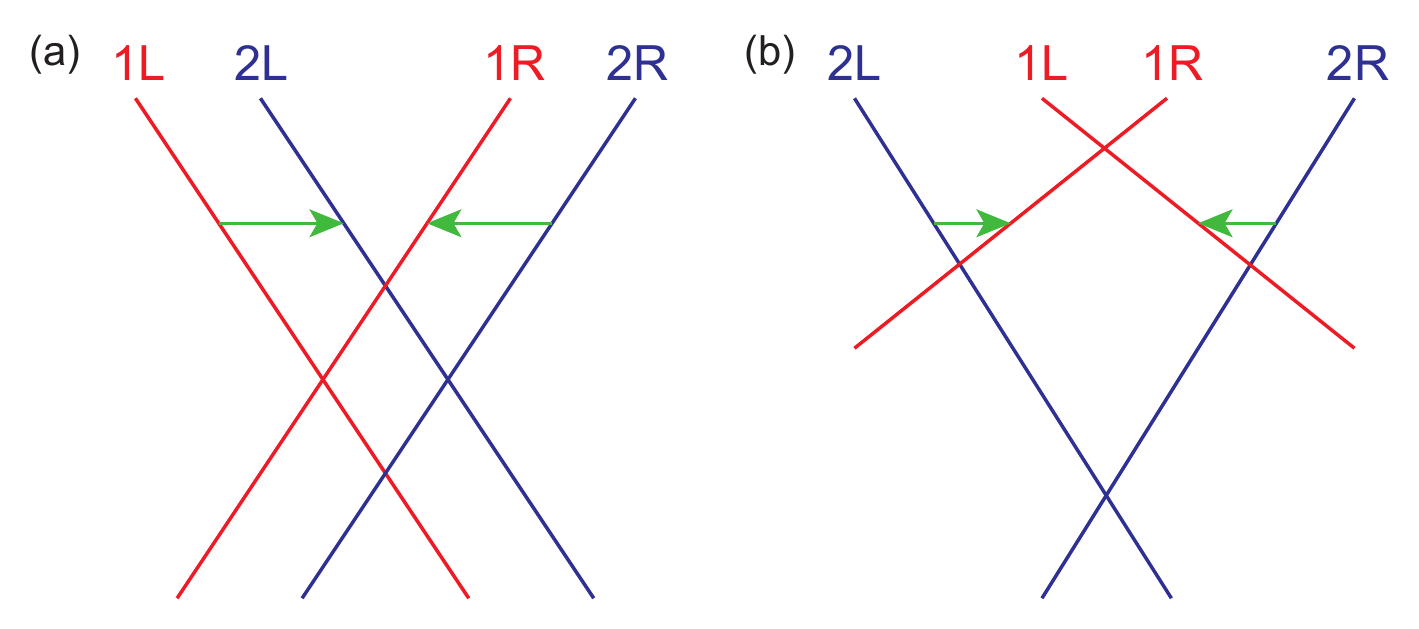}
\caption{Backscattering processes for $n=2$. There are 
two possible cosine terms that represent backscattering: 
(a) $\cos(2\theta_1-2\theta_2)$ and (b) $\cos(2\varphi_1-2\varphi_2)$.
The energy dispersion of (a) might appear when two copies are related by 
time-reversal symmetry. 
When the velocity of two copies are different, the energy dispersion would be like (b). 
}
\label{fig:backscattering}
\end{figure}

Next we consider $\bm{L}_1$ and $\tilde{\bm{L}}_2$. 
$\tilde{\bm{L}}_2$ corresponds to 
\begin{equation}
\tilde{V}_2 = \tilde{g}_b \int dx ( \psi_{1R}^\dagger \psi_{1L}^\dagger \psi_{2L} \psi_{2R} + \text{h.c.} ), 
\end{equation}
and its bosonized form is 
\begin{align}
\tilde{V}_2 &= \frac{2\tilde{g}_b}{(2\pi\alpha)^2} \int dx \cos(2\varphi_1 - 2\varphi_2) \notag \\
&= \frac{2\tilde{g}_b}{(2\pi\alpha)^2} \int dx \cos(2\sqrt{2}\varphi_-).
\end{align}
$\tilde{V}_2$ is equivalent to $V_2$ by the redefinition $\psi_{1L}^\dagger \to \psi_{2L}^\dagger$ and $\psi_{2L}^\dagger \to \psi_{1L}^\dagger$. 
For the redefinition, the velocities of the two modes should be the same. 
When the two velocities are different, $V_2$ and $\tilde{V}_2$ read different scattering processes (Fig.~\ref{fig:backscattering}). 

Finally we extend the analysis to the case where the two velocities are different $v_1 \neq v_2$ as well as $K_1 \neq K_2$. 
In this case, the ``charge'' and ``spin'' degrees are no longer separated. 
Here we concentrate on $\tilde{V}_2$. 
In the RG analysis, the scattering process $\tilde{V}_2$ is relevant for $\delta > 0$, where $\delta$ is a scaling dimension of $\tilde{g}_b$, 
i.e., the coupling constant $\tilde{g}_b$ transforms into $\lambda^{\delta} \tilde{g}_b$ under the scaling $r=(x, \tau) \to \lambda r$. 
The scaling dimension $\delta$ is given by $\delta = 2 + \delta_{\cos}$ with $\delta_{\cos}$ being a scaling dimension of $\cos(2\varphi_1 -2\varphi_2)$. 
Following Ref.~\cite{moroz}, $\delta_{\cos}$ is calculated from the correlator $K(r) = \left< \cos[2\varphi_1(r) -2\varphi_2(r)] \cos[2\varphi_1(0) -2\varphi_2(0)] \right>$ as
\begin{equation} 
K(\lambda r) = \lambda^{\delta_{\cos}} K(r).  
\end{equation}
If we assume an infinitely long system at zero temperature, the Euclidean action after integrating $\theta$ fields is
\begin{equation}
S_\varphi = \frac{1}{2} \int \frac{d\omega}{2\pi} \int \frac{dq}{2\pi} \vec{\varphi}(-q, -\omega)^T L(q, \omega) \vec{\varphi}(q, \omega), 
\end{equation}
where a $2\times 2$ matrix $L(q, \omega)$ is defined as
\begin{equation}
L(q, \omega) = \frac{1}{\pi} (q^2 M_\varphi + \omega^2 M_\theta^{-1}).
\end{equation}
Then the correlator $K(r)$ will be
\begin{equation}
K(r) = \frac{1}{2} e^{4I(r)} 
\end{equation}
with 
\begin{widetext}
\begin{equation}
I(r) = \int \frac{d\omega}{2\pi} \int \frac{dq}{2\pi} (e^{iqx-i\omega\tau} -1) e^{-\alpha|q|} ( L^{-1}_{11}+L^{-1}_{22}-L^{-1}_{12}-L^{-1}_{21} ). 
\end{equation}
To perform the integrations over $q$ and $\omega$, we differentiate $I(r)$ with respect to $x$, and then impose the boundary condition $I(0)=0$. 
Following this procedure, we obtain 
\begin{align}
I(r) = &\frac{B-A\eta_1^2}{4\eta_1(\eta_2^2-\eta_1^2)} \left[ \log\left(\frac{\alpha}{\alpha-ix+\eta_1\tau}\right) + \log\left(\frac{\alpha}{\alpha+ix+\eta_1\tau}\right) \right] \notag \\
& -\frac{B-A\eta_2^2}{4\eta_2(\eta_2^2-\eta_1^2)} \left[ \log\left(\frac{\alpha}{\alpha-ix+\eta_2\tau}\right) + \log\left(\frac{\alpha}{\alpha+ix+\eta_2\tau}\right) \right], 
\end{align}
\end{widetext}
where
\begin{gather}
A = \frac{v_1}{K_1} + \frac{v_2}{K_2} + \frac{1}{\pi}(g'_4 + g'_2), \\
B = (\det M_\theta) \left[ v_1K_1 + v_2K_2 -\frac{1}{\pi}(g'_4 - g'_2) \right], 
\end{gather}
\begin{gather}
\eta_{1,2}^2 = \zeta \mp \sqrt{\zeta^2 - (\det M_\varphi)(\det M_\theta)}, \\
\zeta = \frac{v_1^2+v_2^2}{2} +\frac{1}{(2\pi)^2}(g'^2_4 - g'^2_2). 
\end{gather}
Note that $\eta_{1,2}$ can be regarded as renormalized velocities. 
The scaling dimension $\delta_{\cos}$ becomes
\begin{equation}
\delta_{\cos} = -\frac{A\eta_1\eta_2 + B}{\eta_1\eta_2(\eta_1+\eta_2)},
\end{equation}
and $\tilde{V}_2$ is relevant when $\delta > 0$, i.e., 
\begin{equation}
\frac{A\eta_1\eta_2 + B}{\eta_1\eta_2(\eta_1+\eta_2)} < 2.
\end{equation}
For a simple case where $g'_4 = g'_2 = 0$, $\delta_{\cos}$ reduces to 
\begin{equation}
\delta_{\cos} = - \left( \frac{1}{K_1} + \frac{1}{K_2} \right), 
\end{equation}
and $\tilde{V}_2$ is relevant for 
\begin{equation}
\frac{1}{K_1} + \frac{1}{K_2} < 2.
\end{equation}


\begin{thebibliography}{99}

\bibitem{hsieh}
T. H. Hsieh, H. Lin, J. Liu, W. Duan, A. Bansil, and L. Fu, Nat. Commun. {\bf 3}, 982 (2012).

\bibitem{AndoFu} 
Y. Ando and L. Fu, Annu. Rev. Condens. Matter Phys. {\bf 6}, 361 (2015).


\bibitem{ando}
 Y. Tanaka, Z. Ren, T. Sato, K. Nakayama, S. Souma, T. Takahashi, K. Segawa, and Y. Ando, Nat. Phys. {\bf 8}, 800 (2012). 
 
\bibitem{poland}
P. Dziawa, B. J. Kowalski, K. Dybko, R. Buczko, A. Szczerbakow, M. Szot, E. {\L}usakowska, T. Balasubramanian, B. M. Wojek, M. H. Berntsen, O. Tjernberg, and T. Story, Nat. Matter. {\bf 11}, 1023 (2012). 


\bibitem{hasan}
S.-Y. Xu, C. Liu, N. Alidoust, M. Neupane, D. Qian, I. Belopolski, J. D. Denlinger, Y. J. Wang, H. Lin, L. A. Wray, G. Landolt, B. Slomski, J. H. Dil, A. Marcinkova, E. Morosan, Q. Gibson, R. Sankar, F. C. Chou, R. J. Cava, A. Bansil, and M. Z. Hasan, Nat. Commun. {\bf 3}, 1192 (2012). 

\bibitem{vidya1}
Y. Okada, M. Serbyn, H. Lin, D. Walkup, W. Zhou, C. Dhital, M. Neupane, S. Xu, Y. Wang, R. Sankar, F. Chou, A. Bansil, M. Z Hasan, S.D. Wilson, L. Fu, and V. Madhavan, Science, {\bf 341}, 1496 (2013).

\bibitem{vidya2}
I. Zeljkovic, Y. Okada, M. Serbyn, R. Sankar, D. Walkup, W. Zhou, J. Liu, G. Chang, Y. J. Wang, M. Z. Hasan, F. Chou, H. Lin, A. Bansil, L. Fu, and V. Madhavan, Nat. Mater. {\bf 14}, 318 (2015).


\bibitem{arpes}
B. M. Wojek {\it et al.}, arXiv:1505.03414.

\bibitem{serbyn}
M. Serbyn and L. Fu, Phys. Rev. B {\bf 90}, 035402 (2014).

\bibitem{bernevig}
C. Fang, M. J. Gilbert, and B. A. Bernevig, Phys. Rev. Lett. {\bf 112}, 046801 (2014). 

%
\bibitem{zhang}
F. Zhang, X. Li, J. Feng, C. L. Kane, and E. J. Mele, arXiv:1309.7682. 
%

\bibitem{TangFu}
E. Tang and L. Fu, Nat. Phys. {\bf 10}, 964 (2014).

\bibitem{cha}
J. Shen, Y. Xie, and J. Cha, arXiv:1410.4244. 


\bibitem{gu}
R. Zhong {\it et al.}, Phys. Rev. B {\bf 91}, 195321 (2015). 

\bibitem{vidya3}
I. Zeljkovic, D. Walkup, B. Assaf, K. L Scipioni, R. Sankar, F. Chou, and V. Madhavan, arXiv:1501.01299.



\bibitem{teofukane}
J. C. Y. Teo, L. Fu, and C. L. Kane, Phys. Rev. B {\bf 78}, 045426 (2008).



\bibitem{kitaev}
L. Fidkowski and A. Kitaev,  Phys. Rev. B {\bf 83}, 075103 (2011).


\bibitem{ryu}
S. Ryu and S. C. Zhang, Phys. Rev. B {\bf 85}, 245132 (2012).

\bibitem{yao}
H. Yao and S. Ryu, Phys. Rev. B {\bf 88}, 064507 (2013).

\bibitem{qi}
X.-L. Qi, New J. Phys. {\bf 15}, 065002 (2013).


\bibitem{vishwanath}
L. Fidkowski, X. Chen, and A. Vishwanath, Phys. Rev. X {\bf 3}, 041016 (2013).

\bibitem{wang1}
C. Wang and T. Senthil, Phys. Rev. B {\bf 89}, 195124 (2014).


\bibitem{levin}
Z. C. Gu and M. Levin, Phys. Rev. B {\bf 89}, 201113(R) (2014).

\bibitem{coupledwire}
T. Neupert, C. Chamon, C. Mudry, and R. Thomale, Phys. Rev. B {\bf 90}, 205101 (2014).

\bibitem{schnyder}
A. P. Schnyder, S. Ryu, A. Furusaki, and A. W. W. Ludwig, Phys. Rev. B {\bf 78}, 195125 (2008).






\bibitem{inversion}
B. M. Wojek, P. Dziawa, B. J. Kowalski, A. Szczerbakow, A. M. Black-Schaffer, M. H. Berntsen, T. Balasubramanian, T. Story, and O. Tjernberg, 
Phys. Rev. B {\bf 90}, 161202(R) (2014).

\bibitem{fiete}
M. Kargarian and G. A. Fiete, Phys. Rev. Lett. {\bf 110}, 156403 (2013).

\bibitem{liuhsiehfu}
T. H. Hsieh, J. Liu, and L. Fu, Phys. Rev. B {\bf 90}, 081112 (2014). 

\bibitem{dai}
H. Weng, J. Zhao, Z. Wang, Z. Fang, and X. Dai, Phys. Rev. Lett. {\bf 112}, 016403 (2014).

\bibitem{sun}
M. Ye, J. W. Allen, and K. Sun, arXiv:1307.7191. 


\bibitem{internal}
For any 2D system including multilayers, one can choose single-particle basis states that are either even or odd under the reflection $z\rightarrow -z$. 
In this basis, the mirror symmetry takes the explicit form of an $Z_2$ internal symmetry.   

\bibitem{eval}
Mirror operation is the product of the two-fold rotation $C_2$ and the inversion $P$. 
Since in spin-orbit coupled systems $C_2$ acts on electron's spin in addition to its coordinate, 
we have $P^2 C_2^2 = C_2^2 = -1$. Nonetheless, in the presence of $U(1)$ charge conservation, one can always redefine $M$ by combining $ P C_2$ with 
the $U(1)$ transformation $\psi \rightarrow i \psi, \psi^\dagger \rightarrow - i \psi^\dagger$ to restore the property $M^2=1$, with mirror eigenvalue $\pm 1$.   

\bibitem{liu}
J. Liu, T. H. Hsieh, P. Wei, W. Duan, J. Moodera, and L. Fu, Nat. Mat. {\bf 13}, 178 (2014).

\bibitem{mn1}
E. O. Wrasse and T. M. Schmidt, Nano Lett., {\bf 14}, 5717 (2014). 

\bibitem{mn2}
J. Liu, X. Qian, and L. Fu, {\bf 15}, 2657 (2015).

\bibitem{mn3}
C. Niu, P. M. Buhl, G. Bihlmayer, D. Wortmann, S. Bl\"{u}gel, and Y. Mokrousov
Phys. Rev. B {\bf 91}, 201401(2015).  



\bibitem{lu} 
Y. M. Lu and A. Vishwanath,  Phys. Rev. B {\bf 86}, 125119 (2012). 





\bibitem{CS1} X.-G. Wen, Int. J. Mod. Phys. B {\bf 6}, 1711 (1992). 

\bibitem{CS2} X.-G. Wen and A. Zee, Phys. Rev. B {\bf 46}, 2290 (1992). 

\bibitem{Levin1} M. Levin and A. Stern, Phys. Rev. Lett. {\bf 103}, 196803 (2009). 





%
%
%
%
%
%
%
%

\bibitem{sm} Supplementary Materials. 



\bibitem{CDW} D. Schuricht, F. H. L. Essler, A. Jaefari, and E. Fradkin, Phys. Rev. B {\bf 83}, 035111 (2011). 

\bibitem{chiu}
C.-K. Chiu, H. Yao, and S. Ryu, Phys. Rev. B {\bf 88}, 075142 (2013).

\bibitem{furusaki}
T. Morimoto and A. Furusaki, Phys. Rev. B {\bf 88}, 125129 (2013).

\bibitem{sato}
K. Shiozaki and M. Sato, Phys. Rev. B {\bf 90}, 165114 (2014).

\bibitem{disorder}
A argument of similar spirit has been made in Ref.~\cite{AndoFu} to prove that in the presence of time-reversal symmetry, TCI surface states cannot be localized under disorder.    


\bibitem{wang2}
C. Wang, A. C. Potter, and T. Senthil, Science {\bf 343}, 6171 (2014).

\bibitem{hughes}
M. F. Lapa, J. C. Y. Teo and T. L. Hughes, arXiv:1409.1234. 


\end{thebibliography}

\begin{thebibliography}{9}

\bibitem[49]{bosonization} For bosonization, see for example T. Giamarchi, {\it Quantum Physics in One Dimension} (Oxford University Press, Oxford, 2003). 

\bibitem[50]{millis} S. P. Strong and A. J. Millis, Phys. Rev. Lett. {\bf 69}, 2419 (1992). 

\bibitem[51]{Levin2} M. Levin and A. Stern, Phys. Rev. B {\bf 86}, 115131 (2012). 

\bibitem[52]{moroz} A.V. Moroz, K.V. Samokhin, and C.H.W. Barnes, Phys. Rev. B {\bf 62}, 16900 (2000).

\end{thebibliography}
\end{document}